\documentclass{amsart}
\usepackage{amssymb}

\theoremstyle{plain}

\numberwithin{equation}{section}
\input{tcilatex}

\begin{document}
\title{A Possible Candidate for Dark Matter}
\author{Shi-Hao Chen}
\address{Institute of Theoretical Physics, Northeast Normal University \\
Changchun 130024, China. ~~}
\email{shchen@nenu.edu.cn }
\date{Mar. 23, 2002}
\dedicatory{}
\thanks{}

\begin{abstract}
On the basis of a new quantum field theory without divergence presented by
us we identify W-matter with dark matter and guess that the two new stars $%
RXJ1865$ \ and $3C58$ \ and quasi-stellar objects are the compounds of a
F-celestial body with a W-celestial body with massive enough mass, and some
earthquakes, some tsunamis and some disasters in some areas like Bermuda are
caused by W-objects.
\end{abstract}

\maketitle

\section{ Introduction}

According to present understanding, more than ninety per cent of cosmic
matter is dark matter, dark matter has only gravitational effects and can
hardly be detected unless by gravitation. Many suppositions for dark matter
have presented. Some theories consider that the so-called dark matter may
possibly be composed of yet undetected particles. A common character of the
suppositions is that dark matter are not absolutely dark, i.e., it can be
detected by some way besides gravitation in essence.

In contrast with the suppositions, on the basis a new quantum field theory
without divergence we identify dark matter with the W-particles which cannot
be detected except$\ $gravitation. By the new quantum field theory we have
obtained possible answers to the following problems$^{[1]}$.

1. The issue of the cosmological constant.

2. The problem of divergence of Feynman integrals with loop diagrams.

3. The problem of the origin of asymmetry in the electroweak\ unified theory.

The Lagrangian density of the quantum field theory without divergence is%
\begin{equation*}
\mathcal{L=L}_{F}\mathcal{+L}_{W\text{ }},
\end{equation*}
$\mathcal{L}_{F}$ describes the F-particles which are correspondent to the
particles in the conventional quantum field theory, $\mathcal{L}_{W\text{ \ }%
}$describes the W-particles which are new particles. $\mathcal{L}_{F}$ and $%
\mathcal{L}_{W}$ are symmetric. Thus the F-particles and the W-particles are
fully symmetric. $\mathcal{L}_{F}$ and $\mathcal{L}_{W}$ are independent of
each other before quantization unless the gravitation or repulsion is
considered. After quantization $\mathcal{L}_{F}$ and $\mathcal{L}_{W}$ are
dependent on each other. Thus, there is no interaction between the
W-particles and the F-particles by quantizable fields, only the gravitation
or repulsion may possibly be the interaction between the W-particles and the
F-particles, and a real F-particle cannot transform into a real W-particle
unless by a gravitation field or a repulsion field, and vice versa (but the
two sorts of virtual particles can transform from one into another). The
W-particles form the W-world, and the F-particles form the F-world. Thus we
existing in the F-world cannot detect the W-particles by another methods
except gravitation or repulsion in essence. Both energies of the W-particles
and energies of the F-particles are positive. In present paper we discuss
only this possibility that there is gravitation between W-particles and
F-particles.

In the second section the conjecture that W-matter is identified with dark
matter is presented. In the third section the properties and predictable
astronomical observation of the sort of dark matter are discussed. Section
four is conclusion.

\section{W-matter is identified with dark matter}

Because there are the two sorts of particles corresponding to $\mathcal{L}%
_{W}$ and $\mathcal{L}_{F}$ according to the new quantum field theory$^{[1]}$%
, the energy-momentum tensor $\mathcal{T}_{\mu \nu }$ should be written as 
\begin{equation}
\mathcal{T}_{\mu \nu }=\mathcal{T}_{F\mu \nu }+\mathcal{T}_{W\mu \nu }. 
\tag{1}
\end{equation}%
Correspondingly, the Einstein's equation should also be written as 
\begin{equation}
R_{\mu \nu }-\frac{1}{2}g_{\mu \nu }R+\lambda g_{\mu \nu }=-8\pi G\left( 
\mathcal{T}_{F\mu \nu }+\mathcal{T}_{W\mu \nu }\right) .  \tag{2}
\end{equation}

Because there is no other interaction between the F-particles and the
W-particles except the gravitation, we existing in F-world cannot detect the
W-particles by other methods except the gravition. Thus, if W-particles
exist, they must be the dark matter for the F-world. Because the F-world and
the W-world are symmetric, it is possible that the W-matter is 50 per cent
of all matter in the cosmos. Other components of dark matter for the F-world
may possibly be other undetected F-matter.

In Ref.[1] we obtain naturally a new $SU(2)\times U(1)$ electroweak unified
model whose $\mathcal{L=L}_{F}\mathcal{+L}_{W}$ , here $\mathcal{L}$ is
left-right symmetric and $\mathcal{L}_{W}$ and $\mathcal{L}_{F}$ are both
left-right asymmetic. Thus the world is left-right symmetric in{\Large \ }%
principle, but the part observed by us is asymmetric because $\mathcal{L}%
_{F} $ is asymmetric. This model does not contain any unknown particle with
a massive mass. The world in which we exist, i.e., the F-world, is left-hand
world, then the W-world is the right-hand world. Thus the right-hand world
is the dark matter world for the left-hand world, and vice versa.

\section{The properties and predictable astronomical observation of the sort
of dark matter.}

Because F-particles and W-particles are symmetic, we may suppose that in
some phase after large explosion of the cosmos, the W-elementary particles
and the\ F-elementary particles, e.g. the W-leptons and W-quarks and the
F-leptons and F-quarks, are equal in quantity and mix uniformly. When the
F-elementary particles go together, they can form F-nucleons, F-atoms,
F-molecules, F-matter, F-celestial bodies and the F-world by the strong
interaction and electroweak interaction. Identically, when the W-elementary
particles go together, they can form W-nucleons, W-atoms, W-molecules,
W-matter, W-celestial bodies and the W-world. But when W-elementary
particles and F-elementary particles go together, they cannot form any new
particles, since there is no interaction except the gravitation between
them. Thus, we see that the dark matter for the F-world, i.e., the W-matter,
has the following properties:

1. Dark matter and matter are no longer uniformly mixing.

2. Dark matter can also form a dumpling\ as matter does.

3. Dark matter is absolutely dark, and cannot be detected except by its
gravitation effects.

4. Dark matter or a W-celestial body is transparent for any F-particle,
e.g., a F-photon or a F-electron, can pass through a W-celestial body
without any resistance. Of course, because of gravitation effects, when
F-photons go towards a W-celestial body, they will have a movement to
violet. When F-photons go away from a W-celestial body, they will have a
movement to red. When F-particles, e.g. F-photons, go by a W-celestial body,
their orbits will be winding, i.e., the W-celestial body has lens effect for
the F-particles.

5. Because there is no force offseting the gravitation between a F-celestial
body and a W-celestial body, a F-body can easily go into the interior of a
W-body, and vice versa, and a W-celestial body is a gravitation potential
well for a F-particle or a F-body. Thus, in contrast with impact of two
F-celestial bodies, when a F-celestial body impacts a W-celestial body (dark
matter), the F-celestial body and the W-celestial body will pass through
each other without any friction or dissipative force. Of course, because of
the powerful gravitation of the W-celestial body, some phenomena analogous
to earthquake will occur for the F-celestial body. For example, when a
W-celestial body with a very large mass passes through the earth, the
entrance of the W-celestial body will form a sea and the exit will form land
or a high mountain because of powerful gravitation effects.\ When the
velocity of a F-celestial body relative to a W-celestial body is not large,
the F-celestial body and the W-celestial body will form a vibration system
because of the gravitation\ between them. Obviously, a F-celestial body can
also rotate around a W-celestial body.

From this We guess it is possible that the reason for drift of the land on
the earth is effect of impact of a W-celestial body with a massive mass to
the earth, and the North Pole and the South Pole are respectively the
entrance and the exit of the W-celestial body.

When a W-celestial body with a small mass impacts the earth, it will cause
some earthquakes or tsunamis. The characters of such events are as follows.

A. Gravitation in some areas will be abnormal for a time.

B. All such events will be abrupt since we cannot detect motion of a
W-object.

C. All such events will be accompanied by gathering of seawater and
atmosphere and charges caused by friction, consequencely powerful windstorm,
thunder and lightning, abnormal magnetic field will appear.

From this we guess that some earthquakes, some tsunamis and some disasters
in areas like Bermuda. are caused by some W-celestial bodies.

6. A F-celestial body can coincide with a W-celestial body when their
relative velocity is equal to zero (It is easily proven that according to
quantum mechanics in this case the gravitation potential energy is not
infinite). We call such a celestial body a W-F-celestial body or a W-F-star.
The characters of a W-F-celestial body are as follows.

A. Because the gravitation mass of a W-F-celestial body is a sum of the
W-celestial body and the F-celestial body contained in it and the
W-celestial body is absolutely dark, the mass density of a W-F-celestial
body will be very large, hence the photons radiated by them will have larger
movement to red.

B. Their temperature is not higher since the mass and density of the
F-celestial body contained in it are not still very large.

It is possible that two new stars $RXJ1865$ \ and $3C58$ have been found.
The radius and temperature of $RXJ1865$ are respectively $11.3km$ and $%
7\times 10^{5\circ }C$\ , and the density and temperature of $3C58$ are
respectively 5 times of a neutron-star and $1\times 10^{8\circ }C.$ We guess
that the new stars $RXJ1865$ and $3C58$\ are such two W-F-stars, and are not
so-called quark-stars. The density and temperature of a W-F-star may is
arbitrary since a W-F-star can be composed of a F-celestial body and a
W-celestial body with an arbitearily large mass. But the density and
temperature of a quark-star should be determinate, and no proof show how
much the density and temperature of a quark-star are respectively. Hence $%
RXJ1865$ and $3C58$\ may possiblely be the two W-F-stars. In fact, if the
gravitation mass of a celestial body is abnormally large, the celestial body
is possibly a W-F-celestial body.

Quasi-stellar objects have very large movement to red and radiate very
powerful energy. We guess that quasi-stellar objects are possibly the
W-F-stars. Thus it is possible that their very large movement to red
originates from the very powerful gravitation of the W-celestial body
contained in quasi-stellar objects, the distances from the earth to
quasi-stellar objects are not very long, and the energy emitted by a
quasi-stellar object is not very powerful.

We will discuss the problems in detail in the following paper.

\section{Conclusions}

On the basis of a new quantum field theory without divergence, we identify
W-matter with dark matter and guess that the two new stars $RXJ1865$ \ and $%
3C58$ \ and quasi-stellar objects are the compounds of a F-celestial body
with a W-celestial body with massive enough mass, and some earthquakes, some
tsunamis and some disasters in some areas like Bermuda are caused by
W-objects.

\end{document}